\documentclass[final,5p,times,twocolumn]{elsarticle}

\usepackage{amssymb}
\usepackage{amsmath}
\usepackage{amsthm}
\usepackage{physics}
\usepackage[colorlinks=true,linkcolor=blue,citecolor=blue]{hyperref}
\usepackage{mathrsfs}
\usepackage{appendix}
\usepackage{cleveref}
\usepackage{etex}
\usepackage{graphicx}
\usepackage{floatflt}
\usepackage{amsfonts}
\usepackage{psfrag}
\usepackage{epsf}
\usepackage{slashed}
\usepackage{booktabs}
\usepackage{mathptmx}
\usepackage{latexsym}
\usepackage{subfigure}
\usepackage [latin1]{inputenc}
\def\footnoterule{\kern 1mm \hrule width 7cm \kern 2.2mm}%

\newcommand{\bea}{\begin{eqnarray}}
	\newcommand{\eea}{\end{eqnarray}}
\newcommand{\be}{\begin{equation}}
	\newcommand{\ee}{\end{equation}}
\newcommand{\on}{\operatorname}

\def\C{\mathbb C}

\def\1{{\bf{1}}}

\def \<{\langle}
\def \>{\rangle}
\begin{document}
\begin{frontmatter}
\title{Identifying local unitary equivalence based on reduction of quantum states}
\author{Yanjun Chu$^1$\corref{corr}}
\author{Chenyang Cui$^1$}
\author{Yuhang Xie$^1$}
\author{Mengli Liu$^1$}
\author{Shao-Ming Fei$^2$\corref{con}}

\cortext[corr]{Corresponding author. E-mail: \href{mailto:10100089@vip.henu.edu.cn}{10100089@vip.henu.edu.cn}}
\cortext[con]{Contact author. E-mail: \href{mailto: feishm@cnu.edu.cn}{feishm@cnu.edu.cn}}
\address{1. School of Mathematics and Statistics, Henan University, Kaifeng 475004, China}
\address{2. School of Mathematics Sciences, Capital Normal University, Beijing 100048, China}
\begin{abstract}
Local unitary equivalence is central to entanglement quantification and classification. Identifying the local unitary equivalence remains a formidable challenge. We address this problem for a class of quantum states with one highly degenerate eigenvalue and the rest non-degenerate simple eigenvalues
 that are pivotal to structured resources in quantum resource theory. We introduce a ``reduction" procedure that maps each state to a ``reduced state" by nullifying the highest-multiplicity eigenvalue and prove that the local unitary equivalence of the original states is equivalent to that of their reduced counterparts. For the resulting pure or non-degenerate reduced states, we employ the existing invariants or fixed-point subgroup criteria to establish a complete discrimination framework, although the existing criteria can not directly identify the local unitary equivalence of the original states. We also verify the local unitary equivalence of two families of single-parameterized multipartite mixed states constructed by perturbing absolutely maximally entangled states from distinct combinatorial origins, demonstrating the efficacy and generality of our approach.
\end{abstract}
\end{frontmatter}

\section{INTRODUCTION}

As one of the most striking signatures of quantum physics \cite{P}, quantum entanglement is a kind of fundamental quantum resource that underpins a broad spectrum of quantum information processing tasks including quantum teleportation \cite{BBC,AF}, quantum computation \cite{D, N}, dense coding \cite{BW}, quantum cryptographic protocols \cite{E,DER}, entanglement swapping \cite{ZZHE}, remote state preparation \cite{BDSS,YZG} and many others.

For a bipartite quantum system, the entanglement is invariant under local unitary (LU) transformations. Nevertheless, two quantum states with the same entanglement of formation \cite{BDSW} are not necessarily local unitary equivalent \cite{N}. LU equivalence thus plays a key role in characterizing entanglement \cite{DVC}. Invariants under LU transformations can be used to detect and classify entanglement, as well as to construct entanglement measures \cite{CAF1, CAF2} and separability criteria \cite{CW1, CW2, Ru, ACF}. However, the classification and characterization of quantum states under LU transformations remain a challenging problem, whose solution would give rise to a significant advance in both fundamental understanding and practical exploitation of entanglement as a quantum resource.

A central strategy in classifying quantum states under LU transformations is to determine the complete set of LU invariants. Numerous results and methods have been developed. The authors in Refs.~\cite{Ra, GRB} proposed an approach to compute all LU invariants, although it is generally not operational. For two-qubit systems, Ref.~\cite{MK} provided a complete set of 18 polynomial invariants. The case of three-qubit states was investigated in Refs.~\cite{LPS, S}, while complete sets of LU invariants were constructed for certain families of bipartite mixed states~\cite{AFPY, AFG, AFG2} and a class of tripartite pure states~\cite{ACFW}.

Furthermore, Refs.~\cite{Kr1, Kr2, LLLQ, WLFZ} studied the LU equivalence of multipartite pure states and obtained several efficient criteria. For non-degenerate bipartite mixed states, Ref.~\cite{FJ} gave an operational LU equivalence criterion. Based on the CP decomposition, the authors in \cite{CJ, CJZ} analyzed coefficient tensors of three-qubit states and established a necessary and sufficient condition for the LU equivalence of general tripartite states. 

Ref.~\cite{BR} considered the equivalence of absolutely maximally entangled (AME) states under stochastic local operations and classical communication (SLOCC), introducing verification methods for AME and $k$-uniform states and disproving the conjecture that all AME states are SLOCC equivalent. Since LU equivalence is a special case of SLOCC equivalence, Ref.~\cite{RRKL} explored the LU classification of AME states and constructed a complete set of computable LU invariants from permutations, which distinguishes inequivalent $2$-unitary operators. The approach is generalized to arbitrary multipartite systems with unequal local dimensions.

Many further results on LU classification and characterization can be found in Refs.~\cite{ZZFJL1}--\cite{CZL}. In a broader context, bipartite permutation matrices under local permutation equivalence were classified in Ref.~\cite{YC}. Using local permutation equivalence classes of diagonal states in two qubit and qubit-qutrit systems , Ref.~\cite{CLZ} conjectured that LU equivalence and local permutation equivalence coincide for diagonal quantum states. Ref.~\cite{LCZ} later proved this conjecture for general isospectral bipartite states and presented a simplified criterion for their LU equivalence.

We consider a class of quantum states with one dominant degenerate  eigenvalue and the rest eigenvalues being distinct simple ones, serves as a part of structured resource states in quantum resource theory (QRT), which can be regarded as  extreme points of sets of  absolutely positive partial transpose  states\cite{SC2025, WCS2026}.
For such a pair of quantum states (either bipartite or multipartite), they are  degenerate, thus, the methods in \cite{FJ} and \cite{ZZFJL1} cannot be directly used to determine their LU equivalence. However, we can reduce the eigenvalue with the highest multiplicity to zero under unitary equivalence, thus obtaining a corresponding pair of quantum states. For simplicity, we refer to these states as the reduced states of the original ones, whose eigenvalue of maximal multiplicity is zero. We prove that the LU equivalence of the pair of reduced states coincides exactly with that of the original pair. Within the class we consider, the reduced states fall into two cases. If the reduced states are pure, we can employ the existing methods to determine their LU equivalence, and thereby establish a criterion for the LU equivalence of the original states. If the supports of the reduced states are non-degenerate, we can judge their LU equivalence by computing the LU invariants presented in \cite{ZZFJL1}. In this way, we finally obtain a criterion for the LU equivalence of the original quantum states.

By  our approach, we can also  prove the LU equivalence between two families of single-parameterized multipartite quantum states. Specifically, we leverage the intrinsic relationship between two types of AME states: one type is the AME states constructed in Refs. \cite{BR, RRKL} (which are typical examples of AME states derived from orthogonal Latin squares and related combinatorial structures), the other type is the AME states constructed from orthogonal arrays in Refs. \cite{ZPFZ2,ZPFZ1}. The two families of single-parameterized multipartite quantum states considered are constructed by perturbing these two types of AME states with a single tunable parameter, which preserves the core structural properties of the original AME states. By exploiting the established connection between these two classes of AME states and combining our discrimination method, we systematically verify the LU equivalence of the two families of single-parameterized states. This application not only demonstrates the effectiveness and generality of our approach, but also provides a new perspective in investigating the LU equivalence of multipartite mixed states derived from AME states, which are of great significance in quantum information processing and quantum resource theory.

\section{Preliminaries}\label{sec:basics}
	
Let $H$ be a finite-dimensional Hilbert space. Denote $L(H)$ the set of all linear operators on $H$, $L_+(H)$ the set of positive operators, and $D(H)$ the set of density matrices on  $H$. We equip $L(H)$ with the Hilbert-Schmidt inner product
$\langle M,N\rangle_{HS}=Tr(M^{\dagger}N) $ for all $M,  N \in  L(H)$.
This inner product induces the Frobenius norm (Hilbert-Schmidt norm)
$\|M\|_F=\sqrt{\langle M, M\rangle_{HS}}=\sqrt{Tr(M^{\dagger}M)}$.

Let $H_1, H_2, \cdots, H_t$ be finite-dimensional Hilbert spaces with $\on{dim} H_i=d_i$ and
$\{|e^i_k\rangle_{k=1}^{d_i}\}$ an orthonormal basis of $H_i$, $i=1,2,...,t$.
A general pure state $|\psi\rangle$ on the tensor product space  $\bigotimes_{i=1}^{t}H_i$ takes the form,
$$
|\psi\rangle=\sum_{i_1=1}^{d_1}\cdots \sum_{i_t=1}^{d_t} a_{i_1i_2\cdots i_t }\left|e^1_{i_1}\right\rangle \otimes\left|e^{2}_{i_2}\right\rangle\otimes\cdots\otimes\left|e^{t}_{i_t}\right\rangle,
$$
where $a_{i_1i_2\cdots i_t}\in \C$ are complex coefficients satisfying the normalization condition
$\sum\limits_{i_1, i_2, \cdots, i_t}a_{i_1i_2\cdots i_t}a_{i_1i_2\cdots i_t}^*=1$, where  $*$ denotes complex conjugation.
A multipartite quantum state $\rho$ on $\bigotimes_{i=1}^{t}H_i$ is a positive semi-definite operator with unit trace, admitting the spectral decomposition,
\begin{equation}\label{e1}
\rho=\sum_{i=1}^{d_1\cdots d_t} \lambda_i\left|v_i\right\rangle\left\langle v_i\right|,
\end{equation}
where $\lambda_i$ and  $|v_i\rangle$ are the eigenvalues and corresponding eigenvectors of $\rho$, respectively.
An operator $A$ on $\bigotimes_{i=1}^{t}H_i$ is called tensor decomposable if it can be written as
$A=\otimes_{i=1}^{t} A_i$, where $A_i \in L(H_i)$, denotes a linear operator on $H_i$ for each $i=1,\cdots, t$ \cite{FJ}.

Let $L(\bigotimes_{i=1}^{t}H_i)$ denote the algebra of all linear operators on $\bigotimes_{i=1}^{t}H_i$. For a Hermitian operator  $A\in L(\bigotimes_{i=1}^{t}H_i)$, the commutant of $A$  forms a subalgebra of
$L(\bigotimes_{i=1}^{t}H_i)$. Let $U(\bigotimes_{i=1}^{t}H_i)$ be the unitary group on $\bigotimes_{i=1}^{t}H_i$.
For a Hermitian operator $A$, the set of unitaries commuting with $A$,
\(C_U(A)=\{u\in U(\bigotimes_{i=1}^{t}H_i)|uA=Au\},\)
is called the unitary commutant of $A$, which is a subgroup of $U(\bigotimes_{i=1}^{t}H_i)$.

For any two quantum states $\rho$ and $\rho^{\prime}$ on $\bigotimes_{i=1}^{t}H_i$, we say that $\rho$  and $\rho^{\prime}$ are unitary equivalent  if there exists a unitary operator $U$ on $\bigotimes_{i=1}^{t}H_i$ such that
 $\rho^{\prime}=U\rho U^{\dagger}$. It is a standard result in linear algebra that $\rho$  and $\rho^{\prime}$ are unitary equivalent if and only if they have the same eigenvalues (counting multiplicity).
Furthermore, $\rho$  and $\rho^{\prime}$ are called LU equivalent
if there exist unitary operators $u_i$ on $H_i$ ($i=1,\cdots,t$) such that $\rho^{\prime}=\left(\otimes_{i=1}^{t} u_i\right) \rho\left(\otimes_{i=1}^{t} u_i\right)^{\dagger}.$ If $\rho$ and $\rho^{\prime}$ are LU equivalent, they necessarily share the same set of eigenvalues $\{\lambda_i\}_{i=1}^{d_1\cdots d_t}$.

Let $Q_1$ and $Q_2$ be unitary operators that diagonalize $\rho$ and $\rho^{\prime}$, respectively,
\begin{equation}\label{e2}
\rho=Q_1\Lambda Q_1^{\dagger}, \quad \rho^{\prime}=Q_2\Lambda Q_2^{\dagger},
\end{equation}
where $\Lambda=\operatorname{diag}\left(\lambda_1, \lambda_2, \ldots, \lambda_{d_1\cdots d_t}\right)$.
If $\rho$ and $\rho^{\prime}$ are unitary equivalent via Eq.\eqref{e2}, then
$\rho$ and  $\rho^{\prime}$ are  LU equivalent  if and only if the coset $C_U(\rho) Q_1 Q_2^{\dagger}$ contains a tensor decomposable element (\cite[Lemma 1]{FJ}). Moreover, for  non-degenerate states $\rho$ and $\rho^{\prime}$, they are LU equivalent if and only if  the set of matrices $\{Q_1D Q_2^{\dagger}|D=\on{diag}\left(e^{i \theta_1}, e^{i \theta_2}, \ldots, e^{i \theta_{M N}}\right)\}$
contains  a  tensor decomposable element for some  $\theta_i \in \mathbb{R}$
(\cite[Theorem 1]{FJ} ).

However, for  degenerate states $\rho$ and $\rho^{\prime}$, the orthogonal eigenvectors corresponding to the same eigenvalue are no longer unique. As a consequence, the fixed subgroup $C_U(\rho)$ does not consist solely of diagonal matrices. Although for certain degenerate states in special cases, such as diagonal quantum states(\cite{CLZ,LCZ}), local unitary equivalence can be judged by the rank of the realigned matrix $u^R$ for  $u\in C_U(\rho) Q_1 Q_2^{\dagger}$(\cite{ACFW}), it remains a highly non-trivial problem to determine whether  $C_U(\rho) Q_1 Q_2^{\dagger}$ contains a tensor decomposable unitary element  for general degenerate quantum states. In what follows, we consider a class of quantum states and reduce them to either pure states or non-degenerate  on support set of states. Under this reduction, the original states and the resulting states are related by LU equivalence. This allows us to resolve the original LU equivalence problem for this class of degenerate quantum states.

For the convenience of our subsequent analysis, we recall the invariant theory of LU equivalence for bipartite quantum states \cite{ZZFJL1}. Let $\rho$ be a quantum bipartite states on $H_1 \otimes H_2$ with the spectral decomposition
\begin{equation}\label{e3}
\rho=\sum_{i=1}^{d_1d_2} \lambda_i\left|\phi_i\right\rangle\left\langle \phi_i\right|,
\end{equation}
where \( |\phi_i\rangle \) are the eigenvectors corresponding to the non-zero eigenvalues \( \lambda_i \).
Each eigenvector is a  pure state of the form
\[
|\phi_i\rangle = \sum_{k=1}^{d_1}\sum_{l=1}^{d_2} a_{kl}^i |k\rangle_1 \otimes |l\rangle_2, \quad a_{kl}^i \in \mathbb{C}\]
satisfying
\( \sum_{k=1}^{d_1}\sum_{l=1}^{d_2}  a_{kl}^i (a_{kl}^i)^* = 1.
\)
Define the matrix $A_i=(a^i_{kl})$ for each eigenvector $|\phi_i\rangle$.
If $\rho$ has $d_1d_2-n$ zero eigenvalues, then the following quantities are LU invariants \cite[Theorem 1]{ZZFJL1}:
\begin{equation}\label{e4}
J^s(\rho)=Tr_2(Tr_1\rho^s), ~s=1,2,\cdots, d_1d_2;
\end{equation}
\begin{equation}\label{e5}
Tr[(A_iA_j^{\dagger})(A_kA_l^{\dagger}\cdots(A_hA_p^{\dagger})], ~i,j,k,l,\cdots,h,p=1,2,\cdots,n,
\end{equation}
where $Tr_1$ ($Tr_2$) denotes the partial traces over the first (second) Hilbert space.

Nevertheless, for degenerate states, the eigenvector decompositions given in Eq.(\ref{e3}) are no longer unique. 
Hence, the invariants defined in Eq.\eqref{e5} are generally non-unique. This renders the verification of the invariants Eq.\eqref{e5} highly nontrivial and practically infeasible. To overcome this difficulty, we investigate a transformation that maps two bipartite quantum states into either pure or  non-degenerate on support set of states, so that the aforementioned invariants can be employed to verify the LU equivalence of the original degenerate states.

\section{Reduction of local unitary equivalence for a class of quantum states}\label{sec3}

Denote $D = d_1\cdots d_t$. We say that two Hermitian matrices $A$ and $B$ of order $D$
are local unitary equivalent if there exists a tensor decomposable unitary matrix  $U=u_1\otimes \cdots\otimes u_t$ ($u_i\in \on{U}(d_i)$) such that $B=U^{\dagger}AU$.  Let $\rho$ and $\rho^{\prime}$ be two multipartite quantum states defined on the tensor product space $\bigotimes_{i=1}^{t}H_i$.
Then $\rho$ and $\rho^{\prime}$ are LU equivalent if and only if the Hermitian matrices $\lambda_0\mathbb{I}_{D}-\rho$ and $\lambda_0\mathbb{I}_{D}-\rho^{\prime}$ are LU equivalent for all real number $\lambda_0$,
where $\mathbb{I}_{D}$ stands for the $D$-dimensional identity matrix.

Let $\rho$ and $\rho^{\prime}$ be multipartite states possessing $s$ distinct non-zero eigenvalues satisfying $\lambda_1>\lambda_2>\cdots>\lambda_k\cdots>\lambda_s>0$.  All their eigenvalues are non-degenerate, apart from $\lambda_k$ with degeneracy $D-s+1$. These eigenvalues satisfy the normalization condition $(D-s+1)\lambda_k+\sum\limits_{i=1,i\neq k}^{s}\lambda_i=1$. Therefore, we have

{\em Theorem.}
Let
$ \hat{\rho}=\frac{1}{\lambda_kD-1}(\lambda_k\mathbb{I}_{D}-\rho)$
and
$\hat{\rho^{\prime}}=\frac{1}{\lambda_kD-1}(\lambda_k\mathbb{I}_{D}-\rho^{\prime}).$ Then 
$\rho$ and $\rho^{\prime}$ are LU equivalent if and only if  $\hat{\rho}$ and $\hat{\rho^{\prime}}$ are LU equivalent.

We remark that the requirement for \(\lambda_k\) to be the maximal eigenvalue is imposed by the positivity constraint when using the invariants of  Ref.\cite{ZZFJL1}, and is not a necessary condition solely for establishing local unitary equivalence as stated in the theorem.

Subsequently, we can transform quantum states $\rho$ and $\rho^{\prime}$ into associated states that facilitate the determination of their LU equivalence. We analyze the cases based on the specific physical properties of the quantum states in question.

\subsection{The case of two distinct non-zero eigenvalues}\label{sec3.1}

Let unitary equivalent multipartite states $\rho$ and $\rho^{\prime}$ have two distinct non-vanishing eigenvalues  $\lambda_1>\lambda_2>0$, with corresponding multiplicities $D-1$ and $1$, respectively.
Let $\hat{\rho}=\frac{1}{\lambda_1-\lambda_2}(\lambda_1\mathbb{I}-\rho)$ and  $\hat{\rho^{\prime}}=\frac{1}{\lambda_1-\lambda_2}(\lambda_1\mathbb{I}-\rho^{\prime})$. Then, $\hat{\rho}$ and $\hat{\rho^{\prime}}$ are unitary equivalent pure states, with eigenvalues $1$ and $0$ and respective multiplicities $1$ and $D-1$.  We can reduce the problem of determining the LU equivalence of the degenerate quantum states $\rho$ and $\rho'$ to that of their corresponding pure states $\hat{\rho}$ and $\hat{\rho'}$. Since $\hat{\rho}$ and $\hat{\rho'}$ are pure states, then  $J^s(\hat{\rho})=J^s(\hat{\rho^{\prime}})=1$, so Eq.(\ref{e4}) is automatically satisfied. To  ascertain  whether $\rho$ and $\rho'$ are LU equivalent, we only need to  compute their invariants given by Eq.(\ref{e5}). As for the invariants in Eq.(\ref{e5}), they corresponds to a unique matrix $A_i$(i=1,2), respectively, which makes the verification straightforward. We present several illustrative examples to demonstrate the effectiveness of this proposed approach.

{\em Example 1.} We consider two qutrit states 
$\rho_1$ and $\rho_2$:
\[\rho_1 = \begin{pmatrix}
\frac{7}{75} & 0 & 0 & 0 & -\frac{2}{75} & 0 & 0 & 0 & -\frac{2}{75} \\
0 & \frac{3}{25} & 0 & 0 & 0 & 0 & 0 & 0 & 0 \\
0 & 0 & \frac{3}{25} & 0 & 0 & 0 & 0 & 0 & 0 \\
0 & 0 & 0 & \frac{3}{25} & 0 & 0 & 0 & 0 & 0 \\
-\frac{2}{75} & 0 & 0 & 0 & \frac{7}{75} & 0 & 0 & 0 & -\frac{2}{75} \\
0 & 0 & 0 & 0 & 0 & \frac{3}{25} & 0 & 0 & 0 \\
0 & 0 & 0 & 0 & 0 & 0 & \frac{3}{25} & 0 & 0 \\
0 & 0 & 0 & 0 & 0 & 0 & 0 & \frac{3}{25} & 0 \\
-\frac{2}{75} & 0 & 0 & 0 & -\frac{2}{75} & 0 & 0 & 0 & \frac{7}{75}
\end{pmatrix}\]
and
\[\rho_2 = \begin{pmatrix}
\frac{3}{25} & 0 & 0 & 0 & 0 & 0 & 0 & 0 & 0 \\
0 & \frac{3}{25} & 0 & 0 & 0 & 0 & 0 & 0 & 0 \\
0 & 0 & \frac{7}{75} & 0 & -\frac{2}{75} & 0 & -\frac{2}{75} & 0 & 0 \\
0 & 0 & 0 & \frac{3}{25} & 0 & 0 & 0 & 0 & 0 \\
0 & 0 & -\frac{2}{75} & 0 & \frac{7}{75} & 0 & -\frac{2}{75} & 0 & 0 \\
0 & 0 & 0 & 0 & 0 & \frac{3}{25} & 0 & 0 & 0 \\
0 & 0 & -\frac{2}{75} & 0 & -\frac{2}{75} & 0 & \frac{7}{75} & 0 & 0 \\
0 & 0 & 0 & 0 & 0 & 0 & 0 & \frac{3}{25} & 0 \\
0 & 0 & 0 & 0 & 0 & 0 & 0 & 0 & \frac{3}{25}
\end{pmatrix}.\]
Both $\rho_1$ and $\rho_2$ share the same eigenvalues: $\lambda_1 = \frac{3}{25}$ (with multiplicity $8$) and $\lambda_2 = \frac{1}{25}$ (with multiplicity $1$). Correspondingly we have
\[\hat{\rho}_i = \frac{1}{\lambda_1 - \lambda_2} \left( \lambda_1\mathbb{ I }_9- \rho_i \right), \quad i=1,2.\]
It is straightforward to verify that $\hat{\rho}_1 = |\phi\rangle\langle\phi|$ and
$\hat{\rho}_2 = |\psi\rangle\langle\psi|$ are pure states, where
$|\phi\rangle = \frac{1}{\sqrt{3}}|00\rangle + \frac{1}{\sqrt{3}}|11\rangle + \frac{1}{\sqrt{3}}|22\rangle$ and $|\psi\rangle = \frac{1}{\sqrt{3}}|02\rangle + \frac{1}{\sqrt{3}}|11\rangle + \frac{1}{\sqrt{3}}|20\rangle$.

To determine the LU equivalence of $\hat{\rho}_1$ and $\hat{\rho}_2$, we compute their invariants given by Eqs. (\ref{e4}) and (\ref{e5}). The matrices $A_1$ and $A_2$ corresponding to $\hat{\rho}_1$ and $\hat{\rho}_2$ are recpectively given by
\[\begin{aligned}
A_1 &= \begin{pmatrix}
\frac{1}{\sqrt{3}} & 0 & 0 \\
0 & \frac{1}{\sqrt{3}} & 0 \\
0 & 0 & \frac{1}{\sqrt{3}}
\end{pmatrix}, ~~~
A_2 &= \begin{pmatrix}
0 & 0 & \frac{1}{\sqrt{3}} \\
0 & \frac{1}{\sqrt{3}} & 0 \\
\frac{1}{\sqrt{3}} & 0 & 0
\end{pmatrix}.
\end{aligned}\]
Since
\[\begin{aligned}
A_1 A_1^\dagger &= \begin{pmatrix}
\frac{1}{3} & 0 & 0 \\
0 & \frac{1}{3} & 0 \\
0 & 0 & \frac{1}{3}
\end{pmatrix}, ~~~
A_2 A_2^\dagger &= \begin{pmatrix}
\frac{1}{3} & 0 & 0 \\
0 & \frac{1}{3} & 0 \\
0 & 0 & \frac{1}{3}
\end{pmatrix},
\end{aligned}\]
we have $\text{Tr}\left( A_1 A_1^\dagger \right)^m = \text{Tr}\left( A_2 A_2^\dagger \right)^m = 1$, which implies that $\hat{\rho}_1$ and $\hat{\rho}_2$ are LU equivalent. Therefore, the original states $\rho_1$ and $\rho_2$ are also LU equivalent.

In the aforementioned scenario, if the multiplicities of the two non-vanishing eigenvalues are exchanged, we can also conduct a similar discussion by  {\em Theorem}. For quantum states $\rho$ and $\rho^{\prime}$ with two distinct non-vanishing eigenvalues $\lambda_1 > \lambda_2 > 0$ of multiplicities $1$ and $D-1$,
$\hat{\rho} = \frac{1}{\lambda_2 - \lambda_1} \left( \lambda_2 \mathbb{I}- \rho \right)$ and
$\hat{\rho'} = \frac{1}{\lambda_2 - \lambda_1} \left( \lambda_2 \mathbb{I} - \rho' \right)$ are unitary equivalent, with eigenvalues $1$ and $0$ and corresponding multiplicities $1$ and $D - 1$, respectively. Thus, they are both pure states. By {\em Theorem}, we have that $\rho$ and $\rho^{\prime}$ are LU equivalent if and only if  $\hat{\rho}$ and $\hat{\rho^{\prime}}$ are LU  equivalent. Our method can also be applied to the case of multipartite states.

{\em Example 2.} 
We consider the  two quantum states in three  qubit system.
\[
\begin{aligned}
\rho_1 &=\frac{1}{6}( \ket{000}\bra{000} + \ket{111}\bra{111}) + \frac{1}{18} ( \ket{000}\bra{111} + \ket{111}\bra{000}) \\
&\quad + \frac{1}{9}( \ket{001}\bra{001} + \ket{011}\bra{011} + \ket{100}\bra{100} + \ket{110}\bra{110}\\
&\quad + \ket{101}\bra{101} + \ket{010}\bra{010} )
\end{aligned}
\]
and
\[
\begin{aligned}
\rho_2 &= \frac{1}{6}( \ket{001}\bra{001} + \ket{110}\bra{110} ) - \frac{1}{18}( \ket{001}\bra{110} + \ket{110}\bra{001} ) \\
&\quad + \frac{1}{9}( \ket{000}\bra{000} + \ket{111}\bra{111} + \ket{100}\bra{100} + \ket{011}\bra{011}\\
&\quad+ \ket{101}\bra{101} + \ket{010}\bra{010}).
\end{aligned}
\]
These two states share  eigenvalues: $\lambda_1= \frac{2}{9}$ (simple) and $\lambda_2 = \frac{1}{9}$ (with multiplicity $7$). We construct the associated states $\hat{\rho}_1$ and $\hat{\rho}_2$ as follows:
\[
\begin{aligned}
\hat{\rho}_1 &= \frac{1}{\lambda_2 - \lambda_1} \left( \lambda_2 \mathbb{I}_8 - \rho_1 \right) \\
&= \frac{1}{2}\left( \ket{000}\bra{000} + \ket{111}\bra{111} + \ket{000}\bra{111} + \ket{111}\bra{000} \right)
\end{aligned}
\]
and
\[
\begin{aligned}
\hat{\rho}_2 &= \frac{1}{\lambda_2 - \lambda_1} \left(\lambda_2 \mathbb{I}_8 - \rho_2 \right) \\
&= \frac{1}{2}\left( \ket{001}\bra{001} + \ket{110}\bra{110} - \ket{110}\bra{001} - \ket{001}\bra{110} \right).
\end{aligned}
\]

Following the notations introduced in \cite{N}, we identify that $\hat{\rho}_1 = \ket{GHZ_0^+}\bra{GHZ_0^+}$ and $\hat{\rho}_2 = \ket{GHZ_1^-}\bra{GHZ_1^-}$, which are two Greenberger-Horne-Zeilinger (GHZ) states. Since $\ket{GHZ_0^+} = \left( I_2 \otimes I_2 \otimes \begin{pmatrix} 0 & 1 \\ -1 & 0 \end{pmatrix} \right) \ket{GHZ_1^-}$, it follows that $\hat{\rho}_1$ and $\hat{\rho}_2$ are LU equivalent. Therefore, we conclude that $\rho_1$ and $\rho_2$ are also LU equivalent.

\subsection{The case of multiple distinct non-vanishing eigenvalues}
Let $\rho$ and $\rho^{\prime}$ be two multipartite states with $s$ distinct nonvanishing eigenvalues $\lambda_1>\lambda_2>\cdots>\lambda_s>0$.  All eigenvalues are simple, except for the eigenvalue $\lambda_1$, which has multiplicity $D-s+1$. Then 
$\hat{\rho}$ and $\hat{\rho^{\prime}}$ are unitary equivalent mixed states having non-zero eigenvalues $\frac{\lambda_1-\lambda_2}{N},\cdots,\frac{\lambda_1-\lambda_s}{N} $ with all multiplicities $1$, where $N=\lambda_1D-1$. The projections onto the supports of $\hat{\rho}$ and $\hat{\rho^{\prime}}$, denoted $\on{Supp}(\hat{\rho})$ and $\on{Supp}(\hat{\rho'})$, are both non-degenerate quantum states. 
Although \(\hat{\rho}\) and \(\hat{\rho}'\) are constructed from \(\rho_1\) and \(\rho_2\) via the same affine transformation and possess identical spectra, one cannot conclude \(J^s(\hat{\rho})=J^s(\hat{\rho}')\) defined in Eq.(\ref{e4}) merely from this construction. Instead, the equality \(J^s(\hat{\rho})=J^s(\hat{\rho}')\) holds as a consequence of their local unitary equivalence. It follows that the LU invariants corresponding to Eq.(\ref{e5}) for $\on{Supp}(\hat{\rho})$ and $\on{Supp}(\hat{\rho'})$ coincide exactly with those of $\hat{\rho}$ and $\hat{\rho^{\prime}}$, respectively. Therefore, in light of \cite[Theorem 1]{ZZFJL1}, we can determine whether $\hat{\rho}$ and $\hat{\rho^{\prime}}$ are LU equivalent by computing the LU invariants In Eq.(\ref{e5}) for $\on{Supp}(\hat{\rho})$ and $\on{Supp}(\hat{\rho'})$. By {\em Theorem}, the LU equivalence of $\rho$ and $\rho^{\prime}$ is thus equivalent to that of  $\hat{\rho}$ and $\hat{\rho^{\prime}}$.

By {\em Theorem}, we  reduce the problem of verifying the LU equivalence of the degenerate quantum states $\rho$ and $\rho'$ to that of their associated mixed states $\hat{\rho}$ and $\hat{\rho'}$ by verifying the invariants given by Eqs.\eqref{e5}.  The conditions given by the invariants Eq.\eqref{e5} corresponding to matrices $A_i$ are also verified straightforwardly. We provide an illustrative example to validate the effectiveness of our proposed method.

{\em Example 3.} We consider two qutrit  states 
\[
\rho_1 =\frac{1}{192}
\begin{pmatrix}
24 &\  0 &\ \  0 & \ 0 & \ \ 0&\ \  0&\ \  0 &\ \  0 &\ \ 0\\
\ 0 & 24 & \ \ 0 &\  0 &\ \  0 & \ \ 0 &\ \ 0 &\ \   0 &\ \  0 \\
\ 0 & \ 0 &\  17  &\  0 & -4 & \ \  0 & -7& \ \ 0 &\ \  0 \\
 \ 0 &\  0 &\ \ 0 & 24 &\ \  0 &\ \  0 &\ \  0 &\ \  0 &\ \  0 \\
\ 0& \ 0 & -4 & \ 0 & \ 14 &\ \ 0 & -4 &\ \  0 &\ \ 0 \\
\ 0 &\  0 &\ \ 0 &\  0 &\ \ 0 &\  24&\ \  0 &\ \  0 & \ \ 0 \\
\ 0 & \ 0 & -7 &\  0 & -4 & 0 &\  17\ \ &\ \ 0 &\ \  0 \\
\  0 &\  0 &\  \ 0 &\  0 &\ \ 0 &\  \ 0 & \ \ 0 &\ 24&\ \  0 \\
\  0& \ 0 &\  \ 0 & \ 0 &\ \ 0&\ \  0 &\  \ 0 & \ \  0 & \ 24
\end{pmatrix}
\]
and
\[
\rho_2 =
\frac{1}{192} \begin{pmatrix}
17 & 0 & 0 & 0 & -4& 0 & 0 & 0 & 7 \\
0 & 24 & 0 & 0 & 0 & 0 & 0 & 0 & 0 \\
0 & 0 & 24 & 0 & 0 & 0 & 0 & 0 & 0 \\
0 & 0 & 0 & 24& 0 & 0 & 0 & 0 & 0 \\
-4 & 0 & 0 & 0 & 14 & 0 & 0 & 0 & 4\\
0 & 0 & 0 & 0 & 0 & 24& 0 & 0 & 0 \\
0 & 0 & 0 & 0& 0 & 0 & 24& 0 & 0 \\
0 & 0 & 0 & 0 & 0 & 0 & 0 & 24 & 0 \\
7& 0 & 0 & 0 & 4 & 0 & 0 & 0 & 17
\end{pmatrix}.
\]
Both states share  eigenvalues: $\lambda_1 = \frac{1}{8}$ (with multiplicity $7$), $\lambda_2 = \frac{3}{32}$ (simple) and $\lambda_3 = \frac{1}{32}$ (simple), satisfying $\lambda_1 > \lambda_2 > \lambda_3$. Note that $\lambda_1 - \lambda_2 = \lambda_3$. We construct the associated states $\hat{\rho}_i$ as follows:
\[
\hat{\rho}_i = \frac{1}{2\lambda_1 - \lambda_2 - \lambda_3} \left( \lambda_1 \mathbb{I}_9 - \rho_i \right)= \frac{1}{\lambda_1} \left( \lambda_1 \mathbb{I}_9 - \rho_i \right), \quad i=1,2.
\] For $\rho_1$, we obtain
\[\hat{\rho}_1=
8(\lambda_1 \mathbb{I}_9  - \rho_1 )=
\frac{1}{24}\begin{pmatrix}
0 & 0 & 0 & 0 & 0 & 0 & 0 & 0 & 0 \\
0 & 0 & 0 & 0 & 0 & 0 & 0 & 0 & 0 \\
0 & 0 & 7 & 0 & 4 & 0 & 7 & 0 & 0 \\
0 & 0 & 0 & 0 & 0 & 0 & 0 & 0 & 0 \\
0 & 0 & 4 & 0 & 10 & 0 & 4 & 0 & 0 \\
0 & 0 & 0 & 0 & 0 & 0 & 0 & 0 & 0 \\
0 & 0 & 7 & 0 & 4 & 0 & 7 & 0 & 0 \\
0 & 0 & 0 & 0 & 0 & 0 & 0 & 0 & 0 \\
0 & 0 & 0 & 0 & 0 & 0 & 0 & 0 & 0
\end{pmatrix}.
\]
$\hat{\rho}_1$ can be expressed in its spectral decomposition as $\hat{\rho}_1 = \sum_{i=1}^{2} \eta_i \ket{\phi_i}\bra{\phi_i}$, where $\eta_1 = \frac{3}{4}$, $\eta_2 = \frac{1}{4}$ and
\[
\ket{\phi_1} = \frac{1}{\sqrt{3}} \left( \ket{02} + \ket{11} + \ket{20} \right),\, \ket{\phi_2} = \frac{1}{\sqrt{3}} \left( \ket{02} -2\ket{11} + \ket{20} \right).
\]

For $\rho_2$, we similarly obtain
\[
\hat{\rho}_2
=\frac{1}{24}
\begin{pmatrix}
\ \ 7 & 0 & 0 & 0 &4& 0 & 0 & 0 & -7 \\
\ \ 0 & 0 & 0 & 0 &0& 0 & 0 & 0 & \ \ 0 \\
\ \ 0 & 0 & 0 & 0 &0 & 0 & 0 & 0 &\  \ 0 \\
\ \ 0 & 0 & 0 & 0 &0 & 0 & 0 & 0 &\  \ 0 \\
\ \ 4 & 0 & 0 & 0 &10 & 0 & 0 & 0 & -4 \\
\ \ 0 & 0 & 0 & 0 &0 & 0 & 0 & 0 &\  \ 0 \\
\ \  0 & 0 & 0 & 0 &0 & 0 & 0 & 0 &\ \  0 \\
\ \ 0 & 0 & 0 & 0 &0 & 0 & 0 & 0 &\ \  0 \\
-7 & 0 & 0 & 0 & -4 & 0 & 0 & 0 & 7
\end{pmatrix}.
\]
$\hat{\rho}_2$ has the spectral decomposition $\hat{\rho}_2 = \sum_{i=1}^{2} \eta_i \ket{\psi_i}\bra{\psi_i}$, where
\[
\ket{\psi_1} = \frac{1}{\sqrt{3}} \left( \ket{00} + \ket{11} - \ket{22} \right),\,
\ket{\psi_2} = \frac{1}{\sqrt{3}} \left( \ket{00} - \ket{11} + \ket{22} \right).
\]

The corresponding matrices with respect to $\hat{\rho}_1$ and $\hat{\rho}_2$ are
\[
A^1_1 =
\frac{1}{\sqrt{3}}\begin{pmatrix}
0 & 0 & 1 \\
0 &1& 0 \\
1 & 0 & 0
\end{pmatrix}, ~~
A^1_2 =
\frac{1}{\sqrt{6}} \begin{pmatrix}
0 & 0 & 1\\
0 & -2 & 0 \\
1& 0 & 0
\end{pmatrix},
\]
\[
A^2_1 =
\frac{1}{\sqrt{3}}\begin{pmatrix}
1 & 0 & 0 \\
0 &1 & 0 \\
0 & 0 & -1
\end{pmatrix},~~
A^2_2 =
\frac{1}{\sqrt{6}} \begin{pmatrix}
1& 0 & 0 \\
0 & -2 & 0 \\
0 & 0 &-1
\end{pmatrix}.
\]
The set of matrices corresponding to the invariants Eq.(\ref{e5}) forms a free group generated by $A^1_1 (A^1_1)^\dagger$, $A^1_1 (A^1_2)^\dagger$, $A^1_2 (A^1_1)^\dagger$ and $A^1_2 (A^1_2)^\dagger$.

For $\hat{\rho}_1$, Since \(A_1^1\) and \(A_2^1\) commute and are symmetric,, any invariant in Eq.(\ref{e5}) takes the form
\[
\operatorname{Tr}( A_1^1 )^\alpha (A^1_2)^\beta ) =c_{\alpha,\beta} \operatorname{Tr}\left( \begin{pmatrix} 0 & 0 & 1 \\ 0 & 1 & 0 \\ 1 & 0 & 0 \end{pmatrix}^{\alpha} \begin{pmatrix} 0 & 0 & 1 \\ 0 & -2 & 0 \\ 1 & 0 & 0 \end{pmatrix}^{\beta}\right),
\]
where $c_{\alpha,\beta}=\left( \frac{\sqrt{3}}{3} \right)^{\alpha}\left( \frac{\sqrt{6}}{6} \right)^{\beta} $. 

Note that  $\alpha, \beta$ are non-negative integers such that $\alpha+\beta$ is even. 
If $\beta$ is even, then $\operatorname{Tr}\left( ( A_1^1 )^\alpha (A^1_2)^\beta \right) =c_{\alpha,\beta}(2+2^{\beta})$. If $\beta$ is odd, then $\operatorname{Tr}\left( ( A_1^1 )^\alpha (A^1_2)^\beta \right) = c_{\alpha,\beta}(2-2^{\beta})$.

For $\hat{\rho}_2$, since $A^2_1$ and $A^2_2$ commute and are diagonal, any invariant in Eq.~(\ref{e5}) has the form
\[
\begin{array}{llll}
\operatorname{Tr}\left( (A^2_1)^\alpha (A^2_2)^\beta \right) \\
=c_{\alpha,\beta}\operatorname{Tr}\left( \begin{pmatrix} 1 & 0 & 0 \\ 0 & 1 & 0 \\ 0 & 0 & -1 \end{pmatrix}^{\alpha} 
\begin{pmatrix} 1 & 0 & 0 \\ 0 & -2 & 0 \\ 0 & 0 & -1 \end{pmatrix}^{\beta} \right)
\end{array}.
\]

If $\beta$ is even, $\operatorname{Tr}\left( (A^2_1)^\alpha (A^2_2)^\beta \right) = c_{\alpha,\beta}(2+2^{\beta}).$
If $\beta$ is odd, $\operatorname{Tr}\left( (A^2_1)^\alpha (A^2_2)^\beta \right) = c_{\alpha,\beta}(2-2^{\beta})$. 

Accordingly, the invariants Eq.(\ref{e5}) for $\hat{\rho}_1$ and $\hat{\rho}_2$ coincide for the same non-negative integers $\alpha$ and $ \beta$, i.e., \[\operatorname{Tr}\left( (A^1_1)^\alpha (A^1_2)^\beta \right)=\operatorname{Tr}\left( (A^2_1)^\alpha (A^2_2)^\beta \right).\]
Therefore, $\hat{\rho}_1$ and $\hat{\rho}_2$ are LU equivalent, which implies that the original states $\rho_1$ and $\rho_2$ are also LU equivalent.

\subsection{Applications to absolutely maximally entangled (AME) states}

As an application of the our method presented, we prove the LU equivalence between two families of single-parameterized multipartite quantum states in this section, by leveraging the relationship between the AME states constructed in Refs. \cite{BR,RRKL} and those constructed from orthogonal arrays in Refs. \cite{ZPFZ1, ZPFZ2}.

Consider the following $4$-partite states:
\[
\rho_1(p) = \frac{1-p}{80}\mathbb{I}_{81} - \frac{1-81p}{80}\ket{\Phi}\bra{\Phi}
\]
and
\[
\rho_2(p) = \frac{1-p}{80}\mathbb{I}_{81} - \frac{1-81p}{80}\ket{\Psi}\bra{\Psi},
\]
where $p \in (0,1)$,
\[
\begin{aligned}
\ket{\Phi} &= \frac{1}{3}\bigl( \ket{0000} + \omega\ket{0121} + \omega^2\ket{0212} + \omega^2\ket{1022}
+ \ket{1110}\\
 &\quad+ \omega\ket{1201} + \omega\ket{2011} + \omega^2\ket{2102} + \ket{2220}\bigr),
\end{aligned}
\]
\[
\begin{aligned}
\ket{\Psi} &= \frac{1}{3}\bigl( \ket{0000} + \ket{0111} + \ket{0222} + \ket{1012} + \ket{1120}\\
&\quad + \ket{1201} + \ket{2021} + \ket{2102} + \ket{2210}\bigr)
\end{aligned}
\]
with $\omega = \frac{-1+\sqrt{3}\mathrm{i}}{2}$ being a primitive third root of unity.
Both $\ket{\Phi}$ and $\ket{\Psi}$ are AME states on four qutrit system, i.e., AME(4,3) states \cite{BR, RRKL}. Specifically, $\ket{\Phi}$ is constructed from an orthogonal array $\mathrm{OA}(9,4,3,2)$ \cite{ZPFZ1, ZPFZ2}, whereas $\ket{\Psi}$ is derived from a pair of orthogonal Latin squares \cite{RRKL}.

The associated states $\hat{\rho}_1(p)$ and $\hat{\rho}_2(p)$ corresponding to $\rho_1(p)$ and $\rho_2(p)$ are given by
\[
\hat{\rho}_1(p) = \frac{80}{1-81p}\left( \frac{1-p}{80}\mathbb{I}_{81}- \rho_1(p) \right) = \ket{\Phi}\bra{\Phi},
\]
and
\[
\hat{\rho}_2(p) = \frac{80}{1-81p}\left( \frac{1-p}{80}\mathbb{I}_{81} - \rho_2(p) \right) = \ket{\Psi}\bra{\Psi}.
\]
Notably, there exist unitary operators $u_1, u_2, u_3, u_4 \in U(3)$ such that $\ket{\Psi} = u_1 \otimes u_2 \otimes u_3 \otimes u_4 \ket{\Phi}$ \cite[Theorem 1]{RRKL}. This implies that $\hat{\rho}_1(p)$ and $\hat{\rho}_2(p)$ are LU equivalent for all $p \in (0, \frac{1}{81}) \cup (\frac{1}{81}, 1)$, where $p = \frac{1}{81}$ is excluded to avoid singularity in the denominator.
Consequently, by {\em Theorem}, any two quantum states in the family $\{\rho_1(p), \rho_2(q) \mid p, q \in (0, \frac{1}{81}) \cup (\frac{1}{81}, 1)\}$ are LU equivalent.

As core resources in quantum communication and computation, AME(4,3) states are high-dimensional multipartite maximally entangled states. The LU equivalence of $\rho_1(p)$ and $\rho_2(p)$ allows their mutual transformation via local operations without altering the entanglement resource value. For example, in multipartite quantum key distribution, the easier-to-prepare state based on experimental conditions like local quantum gate availability can be selected, as both are fully equivalent in key generation rate and security, simplifying experimental implementations.

\section{Conclusion}
We have investigated the identification of LU equivalence for a class of quantum states with  one highly degenerate eigenvalue and the rest non-degenerate eigenvalue. Such states are a class of structured resource states in quantum resource theory (QRT) and can be regarded as  extreme points of sets of  absolutely positive partial transpose (APPT) states \cite{SC2025,WCS2026}. For any pair of such states, we set their highest-multiplicity eigenvalue to zero to obtain the ``reduced states". We have proven that the LU equivalence of such original states coincides exactly with that of their reduced states. The reduced states fall into two cases: either pure, then the existing pure-state LU equivalence methods can be applied to determine their equivalence (and thus that of the original states); or non-degenerate of supports of states, then we use the LU invariants in \cite{ZZFJL1}  to judge the LU equivalence, thereby establishing a criterion for the original states. We have also derived the LU equivalence of two families of single-parameterized multipartite states, constructed by perturbing two types of AME states from orthogonal Latin squares \cite{BR, RRKL} and orthogonal arrays \cite{ZPFZ1, ZPFZ2}. Leveraging the connection between these AME states and our method, we have systematically verified their LU equivalence, demonstrating the effectiveness and generality for our approach, and providing a new perspective for multipartite mixed states derived from AME states, which is critical for quantum information processing and QRT.

Our method holds great potential for further generalization. Following this reduction idea under LU equivalence, our current operations may highlight future investigations on discrimination criteria for the LU equivalence applicable to a broader class of quantum states.
 					
\section*{Acknowledgments:}
The first author would like to thank Profs. Naihuan Jing and Zhu-Jun Zheng for their many insightful discussions regarding the local unitary equivalence of quantum states. S.-M. Fei acknowledges financial support from the Specific Research Fund of the Innovation Platform for Academicians of Hainan Province.
\section*{Data availability}
No data were created or analyzed in this study.

\end{document}